\documentclass[twocolumn,showpacs,aps,prl,letterpaper]{revtex4}
\usepackage{graphicx}
\usepackage{dcolumn}
\usepackage{amsmath}
\usepackage{epsfig}
\RequirePackage{xspace}

\usepackage{relsize}
\begin{document}

\title{
\large \bfseries \boldmath Study of the branching ratio of
$\psi(3770)\to D\bar{D}$ in $e^+e^-\to D\bar{D}$ scattering }

\author{Hai-Bo Li}\email{lihb@ihep.ac.cn}  \author{Xiao-Shuai Qin}\email{qinxs@ihep.ac.cn}
\affiliation{Institute of High Energy Physics, P.O.Box 918,
Beijing  100049, China }
\author{Mao-Zhi Yang}\email{yangmz@nankai.edu.cn}
\affiliation{School of Physics, Nankai University, Tianjin 300071,
China}
\date{\today}
\begin{abstract}
Based on the data of BES and Belle, the production of $D\bar{D}$
in the $e^+e^-\to D\bar{D}$ scattering process is studied in this
paper. We analyze the continuum and resonant contributions in the
energy region from 3.7 to 4.4 GeV.  In the $\chi^2$ fit to data,
we obtain the resonance parameters of $\psi(3770)$, the branching
ratio of $\psi(3770)\to D\bar{D}$ decay by confronting the data to
the theoretical formula where both the contributions of the
resonances, continuum and interference effects are included. We
obtain the branching ratio of $\psi(3770)\to D\bar{D}$ decay is
$97.2\%\pm 8.9\%$, as well as the branching ratio of $\psi(4040)$,
$\psi(4160)\to D\bar{D}$ decays.
\end{abstract}

\maketitle

Physics of $e^+e^-$ annihilation at the energy region of $3- 5$
GeV is interesting. It attracts the focus of both experimental and
theoretical studies. The physics at this energy region involves
several well-established vector resonances, $J/\psi$, $\psi(2S)$,
$\psi(3770)$, $\psi(4040)$, $\psi(4160)$, $_{\cdots}$, which are
bound states of quark and antiquark pair $c\bar{c}$. Studying the
production and decays of these resonances can deepen our knowledge
about dynamics of interactions between quarks. The resonance
$\psi(3770)$ has a mass just slightly above the threshold of
$D\bar{D}$ pair production. Its decays evade the suppression of
the Okubo-Zweig-Iizuka (OZI) rule \cite{OZI}. This is consistent
with the fact that the width of $\psi(3770)$ is about 2 orders
larger than those of $J/\psi$ and $\psi(2S)$ \cite{PDG}. The
$J/\psi$ and $\psi(2S)$ can only decay into non-$D\bar{D}$ final
states, which is suppressed according to the OZI rule. It is
believed that $\psi(3770)$ decays dominantly into $D\bar{D}$ pair.
All the well-established non-$D\bar{D}$ decay modes of
$\psi(3770)$ only show up with the branching ratios at the order
of $10^{-3}-10^{-4}$. The measurement of many other decay modes of
$\psi(3770)$ only gives upper bounds, which shows that decay rates
of these non-$D\bar{D}$ decay modes should be smaller than
$10^{-3}$ or $10^{-4}$\cite{PDG,CLEO1}. The sum of all these
well-measured decay rates is at most at the order of several
percent.

  In the $e^+e^-$ collider the properties of the resonance $\psi(3770)$
are measured through the scattering process $e^+e^-\to
\psi(3770)\to f$, where $f$ can be any final states like
$D\bar{D}$ or any other hadrons. The branching ratio of
$\psi(3770)\to D\bar{D}$ and $\psi(3770)\to$non-$D\bar{D}$ can be
derived from the measured scattering cross section of $e^+e^-\to
D\bar{D}$ and $e^+e^-\to \mbox{hadrons}$. Both BES and CLEO-c
Collaborations measured the cross section of $e^+e^-\to D\bar{D}$
at the center-of-mass energy $E_{\mbox{c.m.}}=3773$ MeV several
years ago \cite{BES1,CLEO2}, and their results are in good
agreement with each other.

The CLEO-c Collaboration also measured the cross section of
$e^+e^-\to\psi(3770)\to \mbox{hadrons}$ at $E_{\mbox{c.m.}}=3773$
MeV \cite{CLEO3}. The difference between this and the cross
section of $e^+e^-\to\psi(3770)\to D\bar{D}$ is found to be
$(-0.01\pm 0.08^{+0.41}_{-0.30})$ nb, which indicates that the
decay rate of $\psi(3770)$ to non-$D\bar{D}$ is tiny. However, the
measurement of the BES Collaboration gives that the branching
ratio of $\psi(3770)\to D\bar{D}$ is $(85.5\pm 1.7\pm 5.8)\%$ or
$(83.6\pm 7.3\pm 4.2)\%$, and the decay branching ratio of
$\psi(3770)$ to non-$D\bar{D}$ is $(14.5\pm 1.7\pm 5.8)\%$ or
$(16.4\pm 7.3\pm 4.2)\%$ \cite{BES2,BES3}, which is not consistent
with CLEO-c's measurement.

On one hand, a large branching ratio of $\psi(3770)$ to
non-$D\bar{D}$ contradicts the fact that the sum of the branching
ratios of all the well-established exclusive non-$D\bar{D}$ decays
is not large enough to give such a large decay rate for the
inclusive decay mode. On the other hand, it is difficult to
understand such a large branching ratio for the non-$D\bar{D}$
decays theoretically. Calculations based on the method of QCD
predict that the branching ratios of both exclusive and inclusive
decays of $\psi(3770)$ to non-$D\bar{D}$ final states are very
tiny, the sum of them is at most about $5\%$ \cite{KY,Kuang,HFC}.
Therefore, the decays of $\psi(3770)\to D\bar{D}$ ( or
non-$D\bar{D}$, an equivalent expression for one problem) is still
an unsolved problem for both theory and experiment.

At the $e^+e^-$ collider, the branching ratio of $\psi(3770)\to
D\bar{D}$ can be derived from the measured cross section of
$e^+e^-\to D\bar{D}$. In this work, we reanalyze the experimental
data of the $e^+e^-\to D\bar{D}$ cross section in the
center-of-mass energy region from 3.74 to 4.4 GeV measured by the
BES \cite{BES4} and the Belle \cite{Belle} Collaborations. We will
include not only the contribution of $\psi(3770)$ itself, but also
the contributions of the other resonances with masses near 3770
MeV. We also include the continuum contribution and its
interference effect with the resonances in the energy region 3.74
to 4.4 GeV. We finally find that the continuum contribution can be
explained as an effect of the tail of $\psi(2S)$, whose mass is
about 40 MeV lower than the threshold of $D\bar{D}$ pair
production. By including the contributions of all the resonances
near 3770 MeV and the interference effects, the branching ratio of
$\psi(3770)\to D\bar{D}$ we derived is apparently different from
that of the BES experiment \cite{BES2,BES3}. The contributions of
the resonances with masses below and above that of $\psi(3770)$
are not included in analyzing the data of the $e^+e^-\to D\bar{D}$
cross section by the BES Collaboration \cite{BES2,BES3}. The
effect of these contributions is important for deriving the
branching ratio of $\psi(3770)\to D\bar{D}$ decay.

In Ref.~\cite{yang}, one of us analyzed the data of the $e^+e^-\to
D\bar{D}$ cross section measured by the BES and CLEO-c
Collaborations at $E_{\mbox{c.m.}}=3773$ MeV \cite{BES1,CLEO2}.
The formula describing the $e^+e^-\to D^0\bar{D}^0$ or $D^+D^-$
cross section is derived
\begin{eqnarray}
&&\sigma (e^+e^-\to D^0\bar{D}^0, D^+D^-)
=\frac{\pi}{3}\frac{(s-4m_D^2)^{3/2}}{s^{5/2}}\alpha^2\nonumber\\
&&\times \left|-F_{D\bar{D}}(s)+\sum_{i}\frac{g_{\psi_i
D\bar{D}}Q_cf_{\psi_i}m_{\psi_i}}{s-m_{\psi_i}^2
+im_{\psi_i}\Gamma_{i}}e^{i\phi_i}\right|^2, \label{cross:section}
\end{eqnarray}
where $s=(p_1+p_2)^2$, $p_1$ and $p_2$ are the momenta of $D$ and
$\bar{D}$ mesons, respectively, $m_D$ is the mass of $D^{\pm}$ or
$D^0$, $\bar{D}^0$, $m_{\psi_i}$ the mass of the $i$th resonance,
$\alpha=1/137$ is the electromagnetic fine-structure constant, and
$Q_c=2/3$ is the electric charge of the $c$ quark. The first term
in the absolute-value squared $F_{D\bar{D}}(s)$ describes the
continuum contribution. The second terms in the summation are the
contributions of all the possible resonances, which are described
by the Breit-Wigner form. The $\Gamma_i$'s are the total decay
widths of resonances $\psi_i$'s, and $\phi_i$'s the relevant
phases of the resonance contributions. $g_{\psi_i D\bar{D}}$ is
the coupling of the resonance $\psi_i$ and $D\bar{D}$, which is
defined by
\begin{eqnarray}
&&\langle D(p_1)\bar{D}(p_2)|\psi(p)_i\rangle \nonumber\\
&=&-ig_{\psi_i D\bar{D}}\epsilon^{(\lambda)}\cdot
(p_1-p_2)(2\pi)^4\delta^4(p-p_1-p_2), \label{couple}
\end{eqnarray}
where $\epsilon^{(\lambda)}$ is the polarization vector of
$\psi_i$, and $\lambda$ stands for the polarization state.
$f_{\psi_i}$ is the decay constant of the $i$th resonance
$\psi_i$, which is defined by
\begin{equation}
\langle 0|\bar{c}\gamma_\mu c|\psi_i\rangle
=f_{\psi_i}m_{\psi_i}\epsilon^{(\lambda)}_\mu . \label{fpsi}
\end{equation}
Using the definition of the decay constant, the leptonic decay
width of the resonance $\psi_i$ is
\begin{equation}
\Gamma_{eei}=\frac{4\pi}{3}\frac{Q_c^2\alpha^2f_{\psi_i}^2}{m_{\psi_i}}.
\label{leptonic}
\end{equation}

 With isospin symmetry, the coupling $g_{\psi_i D\bar{D}}$
and the continuum function $F_{D\bar{D}}(s)$ are the same for both
the production of $D^+D^-$ and $D^0\bar{D}^0$. The difference of
the cross sections of $e^+e^-$ to $D^0\bar{D}^0$ and to $D^+D^-$
is caused by the phase-space difference of $D^0\bar{D}^0$ and
$D^+D^-$. With the coupling $g_{\psi_i D\bar{D}}$, the branching
ratio of the vector resonance $\psi_i$ can be obtained
\begin{equation}
{\cal BR}(\psi_i\to D^0\bar{D}^0,\mbox{or}\;
D^+D^-)=\frac{g_{\psi_i
D\bar{D}}^2(m^2_{\psi_i}-4m^2_D)^{3/2}}{48\pi\Gamma_{i}
m_{\psi_i}^2}.
\end{equation}
One can define the branching ratio of $\psi_i\to D\bar{D}$ as the
sum of $\psi_i\to D^0\bar{D}^0$ and $D^+D^-$, and then the
branching ratio of $\psi_i\to D\bar{D}$ is
\begin{eqnarray}
&&{\cal BR}(\psi_i\to D\bar{D})=\nonumber\\  &&\frac{g_{\psi_i
D\bar{D}}^2[(m^2_{\psi_i}-4m^2_{D^0})^{3/2}+(m^2_{\psi_i}-4m^2_{D^\pm})^{3/2}]}{48\pi\Gamma_{i}
m_{\psi_i}^2}. \label{brDD}
\end{eqnarray}

The summed cross section of $e^+e^-\to D^0\bar{D}^0$ and $D^+D^-$
can be obtained from Eq.(\ref{cross:section})
\begin{eqnarray}
&&\sigma (e^+e^-\to D\bar{D})
=\frac{\pi}{3}\frac{(s-4m_{D^0}^2)^{3/2}+(s-4m_{D^\pm}^2)^{3/2}}{s^{5/2}}\nonumber\\
&&\times \alpha^2|-F_{D\bar{D}}(s)+\sum_{i}\frac{g_{\psi_i
D\bar{D}}Q_cf_{\psi_i}m_{\psi_i}}{s-m_{\psi_i}^2
+im_{\psi_i}\Gamma_{i}}e^{i\phi_i}|^2.\label{cs}
\end{eqnarray}
With the expressions of the leptonic decay width and the branching
ratios of $\psi_i\to D\bar{D}$ in Eqs. (\ref{leptonic}) and
(\ref{brDD}), one can reexpress the cross section of $e^+e^-\to
D\bar{D}$ in Eq.(\ref{cs}) as
\begin{widetext}
\begin{eqnarray}
\sigma (e^+e^-\to D\bar{D})
&=&\frac{\pi}{3}\frac{(s-4m_{D^0}^2)^{3/2}+(s-4m_{D^\pm}^2)^{3/2}}{s^{5/2}}\alpha^2\nonumber\\
&&\times |-F_{D\bar{D}}(s)+\sum_{i}\frac{g(m_{\psi_i}^2)}{\alpha}
\frac{6\sqrt{\Gamma_i\Gamma_{eei}{\cal BR}_i}
m_{\psi_i}^{5/2}}{s-m_{\psi_i}^2
+im_{\psi_i}\Gamma_i}e^{i\phi_i}|^2, \label{cross-section2}
\end{eqnarray}
\end{widetext}
where the function $g(x)$ is defined as
\begin{equation}
g(x)=\frac{1}{\sqrt{(x-4m_{D^0}^2)^{3/2}+(x-4m_{D^\pm}^2)^{3/2}}},
\end{equation}
and ${\cal BR}_i$ is the branching ratio of $\psi_i\to D\bar{D}$.

The data of the cross sections of $e^+e^-\to D\bar{D}$ measured by
BES \cite{BES4} and Belle \cite{Belle} are analyzed with the
formula in Eq.(\ref{cross-section2}). The {\it BABAR}
Collaboration \cite{BABAR} has also made similar measurements
before Belle's experiment \cite{Belle}, and their data are
consistent with Belle's, but for simplicity, we only use Belle's
data in our analysis since they have higher statistics. The data
require that the continuum term should be chosen as
\begin{equation}
-F_{D\bar{D}}(s)=\frac{F_0m_{\psi(3770)}^2}{s-a},
\end{equation}
where $F_0$ and $a$ are parameters to be fitted. The value of the
parameter $a$ is found to be approximately the mass squared of
$\psi(2S)$. Therefore the continuum term can be identified as the
virtual contribution of the resonance $\psi(2S)$, whose mass is
below the threshold of $D\bar{D}$ production. Then the continuum
term is chosen to be
\begin{equation}
-F_{D\bar{D}}(s)=\frac{c_0}{s-m_{\psi(2S)}^2
+im_{\psi(2S)}\Gamma_{\psi(2S)} },
\end{equation}
where $c_0$ is the parameter to be fitted.

The other resonances included in analyzing the $e^+e^-\to
D\bar{D}$ scattering cross section are $\psi(3770)$, $G(3900)$,
$\psi(4040)$, and $\psi(4160)$. Both the resonances are
parametrized as Breit-Wigner form except for $G(3900)$, for which
the square root times a phase factor is used according to {\it
BABAR}'s finding \cite{BABAR}. The cross section of $e^+e^-\to
D\bar{D}$ scattering in the energy range from $3.7$ to $4.4$ GeV
can be expressed as
\begin{widetext}
\begin{eqnarray}
&&\sigma (e^+e^-\to D\bar{D})
=\frac{\pi}{3}\frac{(s-4m_{D^0}^2)^{3/2}+(s-4m_{D^\pm}^2)^{3/2}}{s^{5/2}}\alpha^2|\frac{c_0}{s-m_{\psi(2S)}^2
+im_{\psi(2S)}\Gamma_{\psi(2S)} } \nonumber\\
&&+\frac{g(m_{\psi(3770)}^2)}{\alpha}
\frac{6\sqrt{\Gamma_{\psi(3770)}\Gamma_{ee1}{\cal
BR}_1}\;m_{\psi(3770)}^{5/2}}{s-m_{\psi(3770)}^2
+im_{\psi(3770)}\Gamma_{\psi(3770)}}e^{i\phi}
+c_1\sqrt{\frac{1}{\sqrt{2\pi}\sigma_{G(3900)}}e^{-\frac{(\sqrt{s}-M_{G(3900)})^2}{\sigma_{G(3900)^2}}}
} e^{i\phi_1}\nonumber\\
&&+\frac{g(m_{\psi(4040)}^2)}{\alpha}
\frac{6\sqrt{\Gamma_{\psi(4040)}\Gamma_{ee2}{\cal
BR}_2}\;m_{\psi(4040)}^{5/2}}{s-m_{\psi(4040)}^2
+im_{\psi(4040)}\Gamma_{\psi(4040)}}e^{i\phi_2}\nonumber\\
&&+\frac{g(m_{\psi(4160)}^2)}{\alpha}
\frac{6\sqrt{\Gamma_{\psi(4160)}\Gamma_{ee3}{\cal
BR}_3}\;m_{\psi(4160)}^{5/2}}{s-m_{\psi(4160)}^2
+im_{\psi(4160)}\Gamma_{\psi(4160)}}e^{i\phi_3}|^2,
\label{cross-section3}
\end{eqnarray}
\end{widetext}
where $c_0$, $c_1$ and the phases $\phi_i$'s are free parameters
which will be fitted in the $\chi^2$ fit to the experimental data.
In our fitting the values of the following quantities are fixed
and taken from the PDG \cite{PDG}: $m_{\psi(2S)}=3686.09\pm 0.04$
MeV, $\Gamma_{\psi(2S)}=317\pm 9$ keV; the leptonic width of
$\psi(3770)$ is $\Gamma_{ee1}=0.265\pm 0.018$ keV;
$m_{\psi(4040)}=4039\pm 1$ MeV, $\Gamma_{\psi(4040)}=80\pm 10$
MeV, the leptonic width of $\psi(4040)$ is $\Gamma_{ee2}=0.86\pm
0.07$ keV; $m_{\psi(4160)}=4153\pm 3$ MeV,
$\Gamma_{\psi(4160)}=103\pm 8$ MeV, the leptonic width of
$\psi(4160)$ is $\Gamma_{ee3}=0.83\pm 0.07$ keV. The quantities
$m_{\psi(3770)}$, $\Gamma_{\psi(3770)}$, and the branching ratios
of $\psi(3770)\to D\bar{D}$, $\psi(4040)\to D\bar{D}$ and
$\psi(4160)\to D\bar{D}$ are set free and fitted from the data of
BES \cite{BES4} and Belle \cite{Belle}. The values of
$M_{G(3900)}$ and $\sigma_{G(3900)}$ are varied and finally fixed,
which can give the best fit to the data.

Figure \ref{fig1} shows the nominal fit to the data of the cross
section of $e^+e^-\to D\bar{D}$ measured by BES \cite{BES4} and
Belle \cite{Belle}.  The data at 14 different energy points under
the $\psi(3770)$ peak between 3.73 and 3.80 GeV are from BES
experiments by using the $e^+e^-$ scan~\cite{BES4}, while the rest
of the data at 27 different energy points above 3.80 GeV are from
the exclusive initial state radiation (ISR) production of
$D\bar{D}$ events from electron-positron annihilation at a
center-of-mass energy of 10.58 GeV at Belle with an integrated
luminosity of 673 fb$^{-1}$~\cite{Belle}. The data are corrected
by the ISR \cite{BES1,ISR1,ISR2,ISR3}, and are all Born cross
sections. The solid curve is the best fit to the data by using the
formula in Eq.(\ref{cross-section3}). The dashed curve is the
contribution of the resonance $\psi(3770)$, and the dotted curve
is the continuum contribution from the tail of $\psi(2S)$. The
very asymmetric line shape (dashed line in Fig.~\ref{fig1}) is
caused by the phase-space factor in Eq.~(\ref{cross-section3}).
\begin{widetext}
\begin{center}
\begin{figure}[h]
\epsfig{file=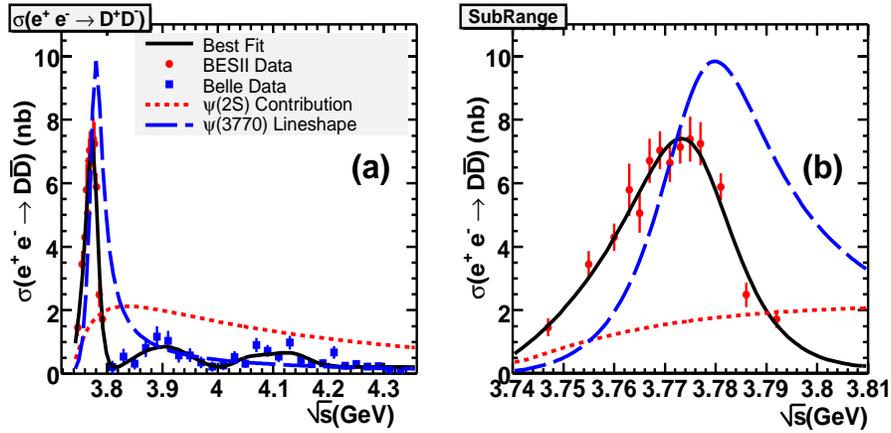,width=12cm,height=6cm} \caption{The cross
section of $e^+e^-\to D\bar{D}$. The dots with error bars are the
data measured by BES \cite{BES4} and squares with error bars are
data from Belle \cite{Belle}. The data are cross sections
corrected by the ISR. The solid curve is the best fit to the data,
the dashed curve is the contribution of the resonance
$\psi(3770)$, and the dotted curve is the continuum contribution
of $\psi(2S)$. (a) is the cross sections in the whole energy range
from $3.7$ to $4.4$ GeV, while (b) is the detail in the region of
the resonance $\psi(3770)$.} \label{fig1}
\end{figure}
\end{center}
\end{widetext}

 The fit gives the following values for the parameters $c_0$,
$c_1$ and the phases:
\begin{eqnarray}
&&c_0=8.75\pm 0.71\;\mbox{GeV}^2,\;\;\; c_1=1.00\pm
0.35\;\mbox{GeV}^{1/2},\nonumber\\
&&\phi=-2.63\pm 0.09,\;\;\; \phi_1=-1.89\pm 0.33,\nonumber\\
&&\phi_2=-2.14\pm 0.14,\;\;\; \phi_3=1.91\pm 0.44,
\label{phase-eq}
\end{eqnarray}
where the unit for $\phi_i$'s is radian.
 The results for the mass and
width of $\psi(3770)$ and the branching ratios of $\psi(3770)$,
$\psi(4040)$, and $\psi(4160)\to D\bar{D}$ are
\begin{eqnarray}
&&m_{\psi(3770)}=3776\pm 1\; \mbox{MeV},\\
&&\Gamma_{\psi(3770)}=28.5\pm 2.1\; \mbox{MeV},
\label{3770-para-eq}
\end{eqnarray}
\begin{eqnarray}
&&{\cal BR}(\psi(3770)\to D\bar{D})=(97.2\pm 8.9)\%, \label{DD-decay}\\
&&{\cal BR}(\psi(4040)\to D\bar{D})=(25.3\pm 4.5)\%, \\
&&{\cal BR}(\psi(4160)\to D\bar{D})=(2.8\pm 1.8)\%,
\label{DD-decay3}
\end{eqnarray}
and the parameters for $G(3900)$ are $M_{G(3900)}=3900\pm 20$ MeV
and $\sigma_{G(3900)}=52\pm 23$ MeV which are fixed in the fit. In
the above fit, there are totally 41 data points measured by BES
and Belle and 11 free parameters are floated as shown in
Eqs.~(\ref{phase-eq})$\sim$ ~(\ref{DD-decay3}). The fitted quality
is $\chi^2/n_d = 1.06$.

In the fit, we find that the correlation between the phase $\phi$
and ${\cal BR}(\psi(3770)\to D\bar{D})$ is 0.73 which is large and
mainly due to the component of the tail of $\psi(2S)$. There is
also a large correlation ($\sim -0.42$) between the parameter
$c_1$ for the structure $G(3900)$ and the ${\cal BR}(\psi(3770)\to
D\bar{D})$. The structure of $G(3900)$ had been suggested in
Ref.~\cite{eichten} and confirmed by {\it BABAR}
data~\cite{BABAR}.

To test the significance of $\psi(2S)$ in the fit, a fit has been
done without the contribution of $\psi(2S)$. We find the quality
$\chi^2/n_d = 1.5$ (the $\chi^2$ in this fit is worse by 14 for 2
degrees of freedom by comparing to the $\chi^2$ value in the
nominal fit) , while the value of $\chi^2/n_d$ is 1.06 in the
nominal fit. Furthermore, in the fit without the contribution of
$\psi(2S)$, we obtain the ${\cal BR}(\psi(3770)\to D\bar{D}) =
(85.9\pm 7.0)\%$, which is significantly smaller than the result
obtained in the nominal fit.

Figure \ref{fig1} shows that the virtual contribution of
$\psi(2S)$ is very large (the dotted curve). It is even larger
than the contribution of $\psi(3770)$ when the colliding energy is
above the resonance region of $\psi(3770)$. The contribution of
any resonance to $e^+e^-\to D\bar{D}$ depends on both the coupling
of this resonance with the virtual photon and with the $D\bar{D}$
pair. The coupling of the resonance with virtual photon can be
described by its decay constant, which can be extracted from the
measured leptonic decay width of the resonance. From the data of
the leptonic width of $\psi(2S)$ and $\psi(3770)$ \cite{PDG}, one
can obtain the decay constant of $\psi(2S)$ is $f_{\psi(2S)}=297$
MeV, while the decay constant of $\psi(3770)$ is
$f_{\psi(3770)}=100$ MeV. This indicates that the coupling of
$\psi(2S)$ with the virtual photon is approximately 3 times of
that of $\psi(3770)$. One can also obtain the coupling of
$\psi(3770)$ and $\psi(2S)$ with $D\bar{D}$ from the fitted result
of ${\cal BR}(\psi(3770)\to D\bar{D})$ and the parameter $c_0$
with identifying
$c_0=g_{\psi(2S)D\bar{D}}Q_cf_{\psi(2S)}m_{\psi(2S)}$, which is
indicated by the numerator of the second term of Eq.(\ref{cs}).
The obtained couplings of the resonances with $D\bar{D}$ are
$g_{\psi(3770)D\bar{D}}=12.8$ and $g_{\psi(2S)D\bar{D}}=12.0$,
i.e., the couplings of $\psi(3770)$ and $\psi(2S)$ with $D\bar{D}$
are approximately the same. Therefore the reason for the large
contribution of $\psi(2S)$ to $e^+e^-\to D\bar{D}$ scattering
comes from the large coupling of $\psi(2S)$ with the virtual
photon.

In Fig. \ref{fig1}, we show one of the best solutions. With the
current data, multisolutions in the fit are possible since both
the four phases and branching ratios are floated in the nominal
fit. We investigate these effects, and find 8 solutions with
comparable fit quality. However 6 of the 8 solutions are with the
${\cal BR}(\psi(3770)\to D\bar{D})$ less than 70\%, which are not
consistent with the fact that no non-$D\bar{D}$ decay modes with
significant branching fraction have been found in experiment. The
sum of the branching ratios of all the well-established
non-$D\bar{D}$ decay modes are at most $2\%$-$3\%$ \cite{PDG}.
Hereafter we discard the nonphysical results. By assuming a
constant width $\Gamma_{\psi(3770)}$ for $\psi(3770)$ resonance in
Eq.~(\ref{cross-section3}), we show two possible physical
solutions for comparison in Table~\ref{tab1:tab}. While, by
assuming an energy-dependent width for $\psi(3770)$ in the fit, we
obtain similar results which are listed in the third column in
Table~\ref{tab1:tab}. The energy-dependent width
$\Gamma_{\psi(3770)}(s)$ is defined as~\cite{BES2}
\begin{eqnarray}
\Gamma_{\psi(3770)}(s) =
\Gamma_{D^+D^-}(s)+\Gamma_{D^0\bar{D}^0}(s)+\Gamma_{non-D\bar{D}}(s),
  \label{width:s:dependent}
\end{eqnarray}
where $\Gamma_{D^+D^-}(s)$, $\Gamma_{D^0\bar{D}^0}(s)$ and
$\Gamma_{non-D\bar{D}}(s)$ are the partial widths for
$\psi(3770)\rightarrow D^+D^-$, $\psi(3770)\rightarrow
D^0\bar{D}^0$ and $\psi(3770)\rightarrow$non-$D\bar{D}$,
respectively, which are taken in the form~\cite{BES2}
\begin{eqnarray}
\Gamma_{D^+D^-}(s) &=& \Gamma_{\psi(3770)} \theta(E_{c.m.} - 2
m_{D^{\pm}})\\ \nonumber &\times &
\left(\frac{p_{D^{\pm}}}{p^0_{D^{\pm}}}
\right)^3\frac{1+(rp^0_{D^{\pm}})^2}{1+(rp_{D^{\pm}})^2} B_{+-},
  \label{width:s:dependent1}
\end{eqnarray}
\begin{eqnarray}
\Gamma_{D^0D^0}(s) &=& \Gamma_{\psi(3770)} \theta(E_{c.m.} - 2
m_{D^0})\\ \nonumber &\times & \left(\frac{p_{D^0}}{p^0_{D^0}}
\right)^3\frac{1+(rp^0_{D^0})^2}{1+(rp_{D^0})^2} B_{00},
  \label{width:s:dependent2}
\end{eqnarray}
and
\begin{eqnarray}
\Gamma_{non-D\bar{D}}(s)=\Gamma_{\psi(3770)}(1-B_{+-}-B_{00}),
  \label{width:s:dependent3}
\end{eqnarray}
where $p^0_D$ and $p_D$ are the momentum of the $D$ mesons
produced at the peak of $\psi(3770)$ and at the center-of-mass
energy $\sqrt{s}$, respectively; $r$ is the interaction radius of
the $c\bar{c}$, which is set to be 1.0 fm here; $B_{+-}$ and
$B_{00}$ are the branching ratios for $\psi(3770) \rightarrow
D^+D^-$ and $D^0\bar{D}^0$, respectively; and $\theta(E_{c.m.} - 2
m_{D^{\pm}})$ and $\theta(E_{c.m.} - 2 m_{D^0})$ are the step
functions to account for the thresholds of $D\bar{D}$ production.
In the fit, we fix the ratio $B_{00}/B_{+-} = 1.33$, so that no
additional free parameter is introduced.

\begin{table}[htp]
\begin{center}
\begin{tabular}{c|c|c|c}\hline\hline
Variables    &  \multicolumn{2}{c|}{Constant width}  & s-dependent   \\
  & Solution 1 & Solution 2 &   \\\hline
$m_{\psi(3770)}$ (MeV)    & 3776$\pm$ 1 & 3776$\pm$1  & 3780$\pm$1  \\
$\Gamma_{\psi(3770)}$(MeV)    &  28.5$\pm$2.1  & 28.7$\pm$2.1  & 29.7$\pm$1.3  \\
${\cal BR}_1$ (\%) &  97.2$\pm$ 8.9  & 101.1$\pm$9.0 & 98.3$\pm$10.4  \\
${\cal BR}_2$ (\%)   & 25.3$\pm$ 4.5    & 34.7$\pm$4.8  & 25.0$\pm$4.6  \\
${\cal BR}_3$ (\%) &  2.8$\pm$ 1.8   & 40.4$\pm$3.8  & 2.9$\pm$1.7  \\
$c_0$ &   8.75$\pm$ 0.71  & 8.67$\pm$0.67  & 10.77$\pm$0.69 \\
$c_1$ & 1.00$\pm$0.35 & 0.82$\pm$0.29 & 1.17$\pm$0.34 \\
$\phi$(rad.) & -2.63$\pm$ 0.09 & -2.56$\pm$0.09  & -2.49$\pm$0.08 \\
$\phi_1$(rad.) & -1.89$\pm$ 0.33 & -1.55$\pm$0.36 & -2.32$\pm$0.30 \\
$\phi_2$(rad.) & -2.14$\pm$ 0.14& -1.62$\pm$0.11 & -2.56$\pm$0.21 \\
$\phi_3$(rad.) & 1.91$\pm$ 0.44  &  -3.03$\pm$0.1 &
1.44$\pm$0.48\\\hline\hline
\end{tabular}
\caption{ Different solutions in the fit to data. The solutions 1
and 2 are two physical solutions with constant width for
$\psi(3770)$ resonance. The last column is one solution for a fit
to data with energy-dependent width ($s$ dependent) as defined in
Eq.~(\ref{width:s:dependent}).} \label{tab1:tab}
\end{center}
\end{table}

Our fitted result for the branching ratio $\psi(3770)\to D\bar{D}$
is different from that of the BES Collaboration \cite{BES2,BES3}.
In our fitting we find that the contribution of the continuum term
from the tail of $\psi(2S)$, the contributions of the resonances
$\psi(4040)$, $\psi(4160)$, the structure $G(3900)$ and their
interference effects with the resonance $\psi(3770)$ are
important; they can seriously affect the fitting result of the
branching ratio of $\psi(3770)\to D\bar{D}$. If these effects are
not included, a smaller result for the decay rate of
$\psi(3770)\to D\bar{D}$ will be obtained. The smaller decay rate
of $\psi(3770)\to D\bar{D}$ means a larger decay rate of
$\psi(3770)\to$ non-$D\bar{D}$. However, no large exclusive
non-$D\bar{D}$ decay mode of $\psi(3770)$ has been seen in
experiment up to now. The sum of all the well-established
exclusive non-$D\bar{D}$ decay rates is less than $2\%$-$3\%$.
This makes a puzzle for $D\bar{D}$ or non-$D\bar{D}$ decays of
$\psi(3770)$. Our new analysis can solve this problem. With the
branching ratio of $\psi(3770)\to D\bar{D}$ is $(97.2\pm 8.9)\%$,
the long-standing problem for $\psi(3770)$ decay disappears. Our
result can also be well understood theoretically. Theoretical
prediction based on QCD to the branching ratio of the
non-$D\bar{D}$ decay of $\psi(3770)$ is at most 5\% \cite{HFC}.
This is consistent with our result of a large branching ratio of
$\psi(3770)\to D\bar{D}$ decay [see Eq.(\ref{DD-decay})].

In summary, we analyze the data of the cross section of $e^+e^-\to
D\bar{D}$ measured by BES and Belle. The data are analyzed by
including the contributions of all the resonances in the energy
region from $3.7$ to $4.4$ GeV. The contributions of the
resonances and their interference effects are important.
Especially the virtual contribution of $\psi(2S)$ is crucial for
obtain our fitting results. Our result for the branching ratio of
$\psi(3770)\to D\bar{D}$ can solve the problem long standing for
$\psi(3770)\to D\bar{D}$ and non-$D\bar{D}$ decays. We also get
the branching ratios of $\psi(4040)$ and $\psi(4160)\to D\bar{D}$.

The authors would like to thank Steve Olsen and Cheng-Ping Shen
for useful discussions and suggestions. This work is supported in
part by the National Natural Science Foundation of China under
Contracts No. 10575108, No. 10735080, No. 10975077, No. 10521003,
No. 10821063, and No. 10835001, the 100 Talents program of CAS,
and the Knowledge Innovation Project of CAS under Contracts No.
U-612 and No. U-530 (IHEP).



\begin{thebibliography}{99}
\bibitem{OZI} S. Okubo, Phys. Lett. 5, 165 (1963); G. Zweig, CERN Rep. 8419/TH-412;
    CERN Preprints TH-401, TH-412; J. Iizuka, Prog. Theor. Phys. Suppl. 37/38, 21 (1966).
\bibitem{PDG} C. Amsler {\it et al.} (Particlr Data Group), Phys.
  Lett. B 667, 1 (2008).
\bibitem{CLEO1}N.E. Adam {\it et al.} (CLEO Collaboration), Phys.
  Rev. Lett. 96, 082004 (2006); T.E. Coan {\it et al.} (CLEO
  Collaboration), Phys. Rev. Lett. 96, 182002 (2006).
\bibitem{BES1}M. Ablikim {\it et al.} (BES Collaboration), Phys.
  Lett. B 603, 130 (2004).
\bibitem{CLEO2}Q. He {\it et al.} (CLEO Collaboration), Phys.
  Rev. Lett. 95, 121801 (2005).
\bibitem{CLEO3}D. Besson {\it et al.} (CLEO Collaboration), Phys.
  Rev. Lett. 96, 092002 (2006).
\bibitem{BES2}M. Ablikim {\it et al.} (BES Collaboration), Phys.
  Lett. B 641, 145 (2006).
\bibitem{BES3}M. Ablikim {\it et al.} (BES Collaboration), Phys.
  Rev. Lett. 97, 121801 (2006).
\bibitem{KY}Y. P. Kuang and T. M. Yan, Phys. Rev. D 41, 155 (1990).
\bibitem{Kuang}Y. P. Kuang, Front. Phys. China 1, 19 (2006).
\bibitem{HFC}Z. G. He, Y. Fan and K. T. Chao, Phys. Rev. Lett. 101, 112001
(2008).
\bibitem{BES4}M. Ablikim {\it et al.} (BES Collaboration), Phys.
  Lett. B 668, 263 (2008).
\bibitem{Belle}G. Pakhlova {\it et al.} (Belle Collaboration), Phys.
  Rev. D 77, 011103(R) (2008).
\bibitem{yang}M.Z. Yang, Mod. Phys. Lett. A 23, 3113 (2008)
 (eprint hep-ph/0610395 )
\bibitem{BABAR}B. Aubert {\it et al.} ({\it BABAR} Collaboration), Phys.
  Rev. D 76, 111105(R) (2007).
\bibitem{ISR1}E.A. Kuraev, V.S. Fadin, Sov. J. Nucl. Phys. 41, 466
(1985).
\bibitem{ISR2} G. Altarelli, G. Martinelli, CERN Yellow
Report 86-02, 47 (1986).
\bibitem{ISR3} O. Nicrosini, L. Trentadue, Phys. Lett. B 196, 551 (1987).

\bibitem{eichten} E. Eichten, K. Gottfried, T. Kinoshita, K. D. Lane, and
T. M. Yan, Phys. Rev. {\bf D 21}, 203 (1980).

\end{thebibliography}
\end{document}